\documentclass[conference]{IEEEtran}
\IEEEoverridecommandlockouts
\usepackage{cite}
\usepackage{amssymb,amsfonts}
\usepackage{algorithmic}
\usepackage{textcomp}
\usepackage{xcolor}
\usepackage{amsmath,graphicx}
\usepackage{booktabs}
\usepackage{tikz}
\usepackage{pgfplots}
\def\BibTeX{{\rm B\kern-.05em{\sc i\kern-.025em b}\kern-.08em
    T\kern-.1667em\lower.7ex\hbox{E}\kern-.125emX}}
\begin{document}

\title{Semi-Supervised Learning Based on Reference Model for Low-resource TTS
}


\author{\IEEEauthorblockN{Xulong Zhang, Jianzong Wang$^\ast$\thanks{$^\ast$Corresponding author: Jianzong Wang, jzwang@188.com.}, Ning Cheng, Jing Xiao}
\IEEEauthorblockA{\textit{Ping An Technology (Shenzhen) Co., Ltd.}}
}

\maketitle

\begin{abstract}
    Most previous neural text-to-speech (TTS) methods are mainly based on supervised learning methods, which means they depend on a large training dataset and hard to achieve comparable performance under low-resource conditions. To address this issue, we propose a semi-supervised learning method for neural TTS in which labeled target data is limited, which can also resolve the problem of exposure bias in the previous auto-regressive models. Specifically, we pre-train the reference model based on Fastspeech2 with much source data, fine-tuned on a limited target dataset. Meanwhile, pseudo labels generated by the original reference model are used to guide the fine-tuned model's training further, achieve a regularization effect, and reduce the overfitting of the fine-tuned model during training on the limited target data. Experimental results show that our proposed semi-supervised learning scheme with limited target data significantly improves the voice quality for test data to achieve naturalness and robustness in speech synthesis.
\end{abstract}


\section{Introduction}

Text-to-speech (TTS) is to covert linguistic features from phonemes to the acoustic features of spectrum to synthesize understandable and natural audio indistinguishable from human recordings. TTS is widely used in application such as voice navigation, telephone banking, voice translation, e-commerce voice customer service, and smart speakers. Generally speaking, most neural TTS methods~\cite{li2019neural,zeng2020aligntts,wang2019semantic,choi2020attentron,shen2018natural,wang2017tacotron,zhao2022nnspeech} utilize two steps to deal with TTS problem. First, they generate mel-spectrogram from the input text information. TTS's primary challenge is the lack of training data. The recording materials of target speakers are pretty limited, which is supposed to be solved urgently. The exposure bias is the main factor for the auto regressive model, it produced by the unmatch between the ground truth data and the generated data. Many existing methods~\cite{schmidt2019generalization,tang2022avqvc,bengio2015scheduled,ranzato2015sequence,wang2022drvc} meet the exposure bias in the module of decoder in the auto regressive model~\cite{juang1985mixture,zhu2019pre}.

The traditional TTS system is mainly build up of two modules, there are front end and back end. The preprocess of text, such as text analysis and language feature extraction, is the main function of the fron end. The back end converts the linguistic features into spectrum of directly raw waveforms. The output is constructed according to the language functions of the front-end and used for speech synthesis. Traditional TTS technology \cite{zen2011product,klatt1987review,asru2021zhang,tokuda2013speech,tokuda2000speech,hunt1996unit,black2007statistical,asru2021tang} is complex and requires professional knowledge in phonetic linguistics.

Neural TTS attracted much attention in the deep learning and speech community in recent years. Most researches use deep neural network-based methods to deal with TTS tasks. WaveNet~\cite{oord2016wavenet} was proposed, this probabilistic auto-regressive model takes linguistic features extracted from input texts as input. While huge data in scale of tens of thounds samples were needed to train the model. Tacotron~\cite{wang2017tacotron} could directly generates waveform signals from input text. The experimental results achieved 3.82 in terms of a mean opinion score (MOS), surpassing production parametric systems in terms of the generated speech's naturalness. 

Shen \textit{et al.} \cite{shen2018natural} proposed Tacotron2 using WaveNet as the vocoder instead of Griffin-Lim \cite{griffin1984signal}, which achieved a MOS of 4.53. Tacotron and Tacotron2 were conditioned on the efficient data, while with limited data, the model works not well. As far as we know, it takes at least ten hours of recording time to build a natural TTS system. Specifically, there are strict needs for the recording environment such as a professional studio for the sound collection. Besides, the content of the sound should cover enough phonemes, and the distribution of the phonemes should be well-tuned. There is very costly and hard to build such a vast and high-quality dataset covered with different speakers. Therefore, it is still a critical task to utilize a few minutes of audio recordings to synthesize any voice in the target's voice, which is to implement TTS under few-shot conditions.

Generally, there will be a degradation in sound quality and robustness when training a TTS model with a limited dataset~\cite{chung2019semi}. To enlarge the capacity of the model for adding new speakers, the pre-trained TTS model was finetuned with the voice of new speaker, which is a research topic name few-shot TTS~\cite{fei2006one,fink2005object}, also known as speaker adaption \cite{choi2020attentron,zhang2020adadurian,sharma2020strawnet,moss2020boffin,chen2018sample,bollepalli2019lombard,deng2018modeling}. However, these methods need a additional process of finetune with the recordings about several minutes or more of the new speaks, and a limited amount of target label data can easily lead to overfitting of the model. Therefore, it has certain limitations: although the process of finetuning cloud change the pretrained model to adapt on new speakers and achieve multi speaker TTS, the training of the model with few samples on the target speaker may lead to error for the cross lingual speaking.

In this paper, we focus on the study of semi-supervised learning scheme, the semi-supervised learning based on the reference model for few-shot neural TTS, which performs well for the inference on out of domain samples. In the method, the reference model based on the backbone network of Fastspeech2 is pre-trained by multiple speakers' amount of recordings. Then the reference model is transferred into the low-data target speaker datasets to be fine-tuned. Meanwhile, pseudo labels generated by the original reference model are used to guide the fine-tuned model's training further, achieve a regularization effect, and reduce the overfitting of the fine-tuned model during training on the limited target data.

\section{Related Works}
\label{sec:related works}
\subsection{Knowledge Distillation}
Knowledge distillation (KD)~\cite{hinton2015distilling} can make student modle get the information from the teacher model. Its success is usually attributed to the privileged information of similarity between the class distribution of the teacher model and the student model. It was first proposed by Hinton \textit{et al.}~\cite{hinton2015distilling} transfer knowledge from large teacher networks to smaller student networks. It works by training students to predict target classification labels and imitate teachers' class probabilities because these features contain additional information about how teachers generalize~\cite{hinton2015distilling}. 

Liu~\textit{et al.}~\cite{liu2020teacher} tried the method of the teacher-student model for resolving the problem of exposure bias. There is an existing problem of exposure bias of autoregressive, due to the unmatched training and inference phase. This problem cloud leads to an unpredictable error for the model during the inference and accumulates the error frame by frame along the time axis.

\subsection{Pseudo Label}

Pseudo labels~\cite{lee2013pseudo} are the predicted labels by the model with the maximum probability for the unlabeled data sample, which may not be the real target class. The pseudo label can alleviate the handcrafted label by the human. During the training phase, the pseudo labels and labels are applied to train the new model in a supervised mode. For unlabeled data, each weight update recalculates the pseudo label, which is used to supervise the model trainging task with the same loss function. Due to the number of the different data have huge different in the data scale, the balance of the different data are very important to the perfromance of the final trained modle.

Higuchi \textit{et al.}~\cite{higuchi2021advancing} tried the method of using pseudo labels to do the automatic speech recognition (ASR), and the results show an improvement with the use of text generated from untranscribed audio. While for the task of TTS, it has often been treated as supervised mode.  A semi-supervised learning method based on the generated label could release the cost of paired data for training.

\section{Proposed Method}
\begin{figure}[htb]
  \centering
 \includegraphics[width=1\linewidth]{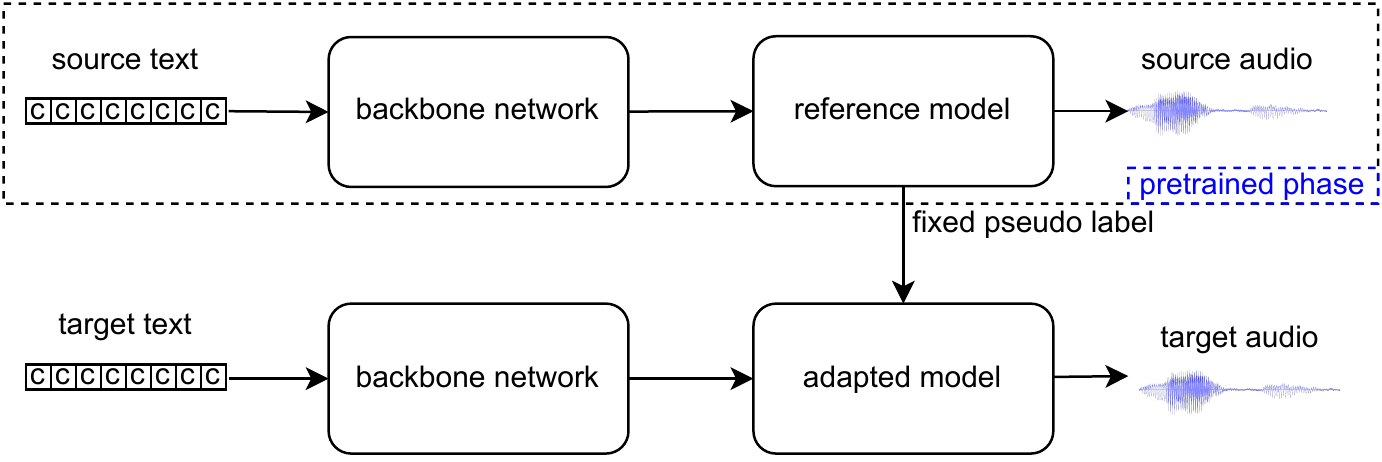}

\caption{Diagram of the semi-supervised learning method based on backbone network in 2 steps: Step 1, pre-training the reference model with abundant source data; Step 2, fine-tuning the original reference model with a limited target dataset, meanwhile pseudo labels generated by the original reference model are used to further guide the training of the adapted model.}
\label{fig:1}
\end{figure}

Our method is a semi-supervised learning scheme. This method works well when the label data is not abundant, and it can address the problem of exposure bias caused by the different processes during autoregressive mode between the inference and training phases. Firstly, we pre-train a backbone network based on Fastspeech2 to introduce a reference loss. The total loss is obtained by configuring the appropriate trade-off parameter $\omega$, where the reference model is fixed during the training iteration process and the fine-tuned model is fine-tuned based on a copied initial reference model.  We illustrate the overall architecture of the proposed semi-supervised learning scheme in Figure~\ref{fig:1}.


The model was mainly built up by hard loss and reference loss. The hard loss is the MSE loss between the mel-spectrogram predicted by the self-training model and the ground truth spectrum. The reference loss is the MSE loss between the predicted spectrum of the pre-trained reference model and the predicted spectrum by the self-training model.

\begin{figure*}[htbp]
	\centering
   \includegraphics[width=1\linewidth]{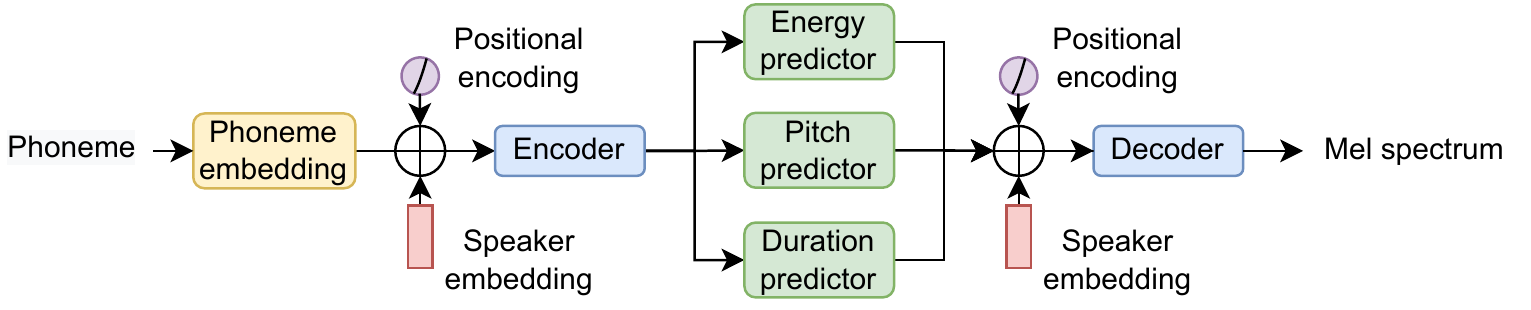}
  
  \caption{The block diagram of the backbone network based on Fastspeech2 has four modules, an encoder module, a decoder module based on feed forward tansformer, predictor and the speaker embedding module.}
  \label{fig:2}
  \end{figure*}
\subsection{The Backbone Network}

We follow the architecture of the main components of Fastspeech2, we use the feed-forward transformer to build up our model. This method uses the network structure of the encoder and decoder in the Fastspeech2 \cite{ren2020fastspeech} model. It is a sequence-to-sequence cyclic feature prediction network, where the encoder converts the input of phoneme sequence into a latent vector, and the decoder is used to predict the output of the mel-spectrogram from the latent vector of the linguistic feature. The vocoder of HiFi-GAN \cite{kong2020hifi} was used for the audio generation from the mel-spectrum. With using waveform generation technology will not affect the validity of the proposed training scheme. 

Figure~\ref{fig:2} shows the overall network structure of the model. The input of this model is a text sequence of the speaker in the training set. After mapping it into a learned 512-dimensional phoneme embedding, it is passed into a encoder with three feed forward transformer modules, the positional encoding and speaker embedding were added to the input of encoder. After the encoder, a fixed-length context vector is obtained. The latent vector was feed to three predictor to predict energy, pitch and duration separtely. With the predicted energy, pitch and duration, the speaker embedding and positional encoding are feed into the decoder for the mel spectrum decoder. The decoder consists of four layers feed forward transformer. The decoder generates the mel-spectrogram from the encoded input text sequence. The speaker embedding module used the X-vector for the speaker representation, it was added both in the encoder and decoder.

\subsection{Semi-Supervised Learning Based on Reference Model}

Machine learning mainly has three categories for the view of lables, there are super vised learning which have all the lables, and unsupervised learning which have no information of the labels, and semi-supervised learning that have partial samples with the labels and left many portions have no labels. Generally speaking, semi-supervised learning trains the model to use both a few labeled data and large amounts of unlabeled data.  In this paper, we use semi-supervised learning to implement a semi-supervised learning method for neural TTS in which labeled target data is limited.

This idea is through using a reference model, which has been trained in advance on a source dataset with sufficient data. Fastspeech2 was chosen for the backbone of our model, because of the high performance of the transformer architecture. The fine-tuned model is generated by the initial reference model on a target dataset with a limited amount of data. At the same time, during the fine-tuning model's continuously training, another initial reference model is fixed to generate pseudo labels to guide the training of the fine-tuned model. As is shown in Figure~\ref{fig:1}, the proposed semi-supervised learning scheme is mainly realized by introducing a reference loss to achieve the MSE regularization effect and reduce the overfitting of the fine-tuned model during training. The total loss is composed of two parts: the hard loss and the reference loss. 

The hard loss is the MSE loss between the fine-tuned model's mel-spectrogram output and the ground truth one. Our goal is to make the output of the fine-tuned model as close as possible to the real acoustic feature, so The hard loss is our main optimization goal and it can be computed as 

\begin{equation}
\mathcal{L}_{\text {hard}}=\frac{1}{N} \sum_{j=1}^{J}\left\|\mathcal{M}_{f t}^{(j)}-\widehat{\mathcal{M}}^{(j)}\right\|^{2}
\label{eq1}
\end{equation}

\noindent where ${\mathcal{M}_{ft}^{(j)}}$ is the mel-spectrogram infered by the fine-tuned model, and ${\widehat{\mathcal{M}}^{(j)}}$ is the natural mel-spectrogram from the ground truth waveform signals. Additionally, ${N}$ represents the number of samples.

The reference loss is the MSE loss between the mel-spectrogram output predicted by the fine-tuned model and the mel-spectrogram output predicted by the initial reference model. We use this loss to get benefit from the trained reference model so that the fine-tuned model can easily achieve a satisfactory adaptive result with a limited target result. The reference loss can be computed as  

\begin{equation}
\mathcal{L}_{\text {ref}}=\frac{1}{N} \sum_{j=1}^{J}\left\|\mathcal{{M}}_{ft}^{(j)}-\mathcal{M}_{ref }^{(j)}\right\|^{2}
\label{eq2}
\end{equation}

\noindent where ${\mathcal{M}_{ref}^{(j)}}$ is the mel-spectrogram inferred by the reference model. Then the total loss can be computed as 

\begin{equation}
\mathcal{L}_{\text {total}}=\mathcal{L}_{\text {hard}}+\omega \mathcal{L}_{\text {ref}}
\label{eq3}
\end{equation}

\noindent where $\omega$ is a trade-off parameter for the two terms. The total loss is obtained by configuring the appropriate $\omega$. If $\omega$ is too high, the fine-tuning model will pay too much attention to the output of the initial reference model. If $\omega$ is too small, the performance of the source data will be lost. The reference loss aims to keep the fine-tuning model to a certain degree of generalization, while the hard loss is to maintain the target person's tone. 

In a sense, this method is aimed at the few-shot situation. The semi-supervised learning scheme based on the reference model solves the problem that the target dataset is insufficient. According to Eq. (\ref{eq2}), the reference loss inspired \cite{lee2013pseudo,xie2020self,hinton2015distilling} is in effect equivalent to MSE regularization. By minimizing both the hard loss and the reference loss, we can minimize the total loss to realize a better generalization performance. Due to the insufficient amount of label target data, pseudo labels generated by the reference model for this small amount of target data. Therefore, to some extent, the introduction of pseudo labels solves the problem of insufficient label target data. Compared with only fine-tuning the reference model, it provides a large amount of pseudo labels data, avoiding limited data. The overfitting problem caused by training the fine-tuned model alone on the limited target data has also been solved.

\section{Experiments and Results}

\subsection{Dataset}

The dataset consists of recordings from three different Chinese native speakers. The recordings of speaker one and speaker two are the labeled source dataset to pre-train the reference model, which contains a total of 24100 recordings. Speaker three, as the target labeled dataset to train the fine-tuned model, has 8915 recordings. We randomly choose 100 recordings of speaker three for the use of test data and left for the training data.

\subsection{Training Setup}

We trained our model on a single Tesla V100 PCLE 16GB GPU. The batch size of each training set is 16. Simultaneously, we use a gradient clip threshold of 1.0 to avoid gradient exploding problems. We use the same learning rules described above to train the reference model on the sufficient source dataset. In our experiments, an 80-channel mel-spectrogram is generated by the encoder from the input text sequence. The trade-off parameter $\omega$ is set as 0, 0.1, 0.5, and 1.0 for the training of the different data size target dataset.

\subsection{Subjective Evaluation}

We conducted the evaluation of robustness and naturalness of the fine-tuned models on the test dataset, which has no overlap with the training dataset. We select 40 native Chinese speakers, including 18 men and 22 women, to participate in the subject evaluation tests. Each person listens 210 audio samples in totally consisted groudtruth recordings and audio generated by the model corresponding to 20 fine-tuned models, and each model is tested with 1000 recordings.

\subsubsection{Naturalness Evaluation}

To evaluate the proposed model, we coducted a MOS test for the whole quality of naturalness. The subjects rate the quality on a range of  5-point scale, the higher the better. Different subjects are independent during the evaluation, and the scores of different models given by the testers are not directly compared.

\begin{figure}[htb]
    \centering
    \begin{tikzpicture}
 \begin{axis}[width=9cm, height=7.2cm, tick align=outside,legend pos=south east,]
   \addplot[draw=red, mark=star, line width=1.05] coordinates {(30,3.18) (300,3.60) (500,3.96) (1000,4.06) (2000,4.14)};
   \addlegendentry{$\omega$=0.1}
   \addplot[draw=orange, mark=diamond, line width=1.05] coordinates {(30,3.34) (300,3.52) (500,3.69) (1000,3.74) (2000,3.81)};
   \addlegendentry{$\omega$=1.0}
   \addplot[draw=blue, mark=triangle, line width=1.05] coordinates {(30,2.62) (300,3.4) (500,3.26) (1000,3.38) (2000,3.52)};
   \addlegendentry{$\omega$=0.5}
   \addplot[draw=green, mark=square, line width=1.05] coordinates {(30,2.7) (300,3.48) (500,3.62) (1000,3.86) (2000,4.16)};
   \addlegendentry{$\omega$=0(baseline)}
 \end{axis}
 \end{tikzpicture}
    \caption{Variation of MOS with the given audio samples number, for 20 different models corresponding to the target data volume of different recordings (30, 300, 500, 1000, 2000) and different trade-off parameters $\omega$ (0, 0.1, 0.5, 1.0).}
    \label{fig:3}
\end{figure}
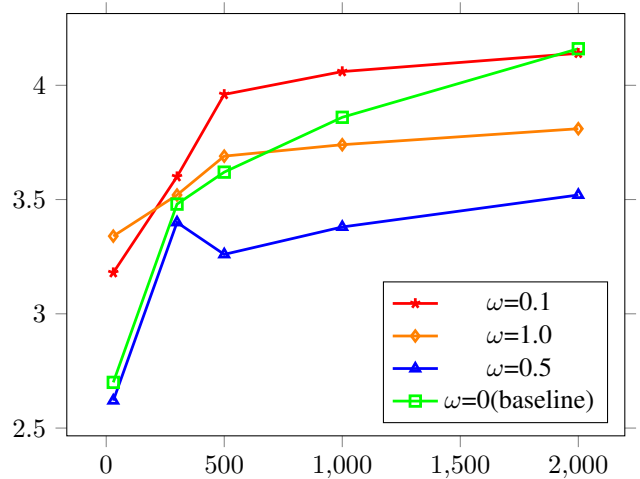

In order to show our results more intuitively, we present the results in the form of a line chart. As shown in Figure~\ref{fig:3}, variation of MOS with the given audio samples number, for 20 different models corresponding to the target data volume of different recordings (30, 300, 500, 1000, 2000) and different trade-off parameters $\omega$ (0, 0.1, 0.5, 1.0). Without informing the subjects, we also conducted a MOS evaluation of the natural speech recordings provided this time compared to the experimental results.

According to the experimental results, we can see from the changing curve corresponding to different trade-off parameters $\omega$. When $\omega$ is 0.1, and the target data size is small (recording = 30, 300, 500, 1000), the MOS value close to or far exceeds the baseline ($\omega = 0$), so the fine-tuned model's generalization ability can be achieved to a certain extent by introducing the reference loss and setting reasonable trade-off parameters. Nevertheless, when we look at other different trade-off parameters ($\omega = 0.5, 1.0$), in most cases, the performance of the baseline is not exceeded, so we can infer that the possible reason is that the value of $\omega$ is too large. When $\omega$ is too large, the fine-tune method we proposed will be forced to output mel-spectrum as close as that of the initial reference model. Excessive penalty leads to underfitting problems in the fine-tuned. The MOS value of the trade-off parameter $\omega$ (0.1 and baseline) corresponding to different recordings can be obtained in Table~\ref{table:1}.
 
\subsubsection{Robustness Evaluation}

We conduct further experiments on these fine-tuned models for speech synthesis. We evaluate the robustness by Word Error Rate (WER $\%$), which evaluate the robustness through the WER, the sum of repetitions (insertions), and the skips (deletions) in this auditory test in this test set. Repeat and skip represent two types of errors for the fine-tuned model in this paper. It shows that the model can run this WER effectively in Table~\ref{table:1}.

\begin{table}[th]
\centering
\caption{Numerical comparison of models for different dataset sizes.}
\label{table:1}
\begin{tabular}{cccc}
\hline
\textbf{Dataset size}&\textbf{$\omega$}& \textbf{MOS}&\textbf{WER $\%$}\\
\hline
30& 0 (baseline)   &  2.70 ± 0.04   &       3.0 \\
300& 0 (baseline)  &  3.48 ± 0.05   &       1.5 \\
500& 0 (baseline)  &  3.62 ± 0.03   &       0.5 \\
1000& 0 (baseline) &  3.86 ± 0.04   &       1.0\\
2000& 0 (baseline) &  4.16 ± 0.03   &       0.5\\
30& 0.1            &  3.18 ± 0.04   &       1.5\\
300& 0.1           &  3.60 ± 0.05   &       1.0 \\
500& 0.1           &  3.96 ± 0.04  &  0.2\\
1000& 0.1          &  4.06 ± 0.04   &       0.2\\
2000& 0.1          &  4.14 ± 0.02  &       0.2\\
\hline
\end{tabular}
\end{table}

Then, looking at the results, whether from the figure or the results table, we can find an interesting phenomenon, firstly, for our model, when $\omega$ is 0.1 and the number of samples is 2000, the best performance is got, but if we observe it more carefully, we can easily find that with 30 training sentences $\omega$ = 1.0 is slightly better than $\omega$ = 0.1, which are both significantly better than $\omega$ = 0.5 and the baseline. but then add 270 sentences and that is all turned on its head, If another 200 sentences are added, then the proposed regularisation apparently gives better performance than the baseline again. which indicates that how to choose the size of $\omega$ is very important for the performance of the model and the selection of $\omega$ is closely related to the sample size. 

Besides, as is shown in Figure~\ref{fig:3}, when the number of samples increase the MOS also increase. When the number of samples increases to a certain extent, the reference model will have a little effect on the learning process of our fine-tune model, it nearly same as the finetune of the backbone only. Specifically, when the number of samples lower than 1000, the reference loss encourages the output of our model to be similar to that of the reference model, which will limit the ability of our model to learn better output.

What's more, with the number of samples being 2000, the performance of baseline has exceeded our best model($\omega$ = 0.1 and the sample number is 500), this seemingly bad result just shows the effectiveness of our method. As is noted, our model is proposed to issue the problem of low resource TTS, In other words, we do not need to consider the case that the number of samples is more than 2000 because so much training data is not need in the case of low resource adaptive. While when the number of the training sample is less than 2000, our model always outperforms baseline, this phoneme determines that the method we proposed works. 

\subsection{Analysis}
\begin{figure}[htb]
\begin{minipage}[b]{1.0\linewidth}
  \centering
  \centerline{\includegraphics[width=9cm]{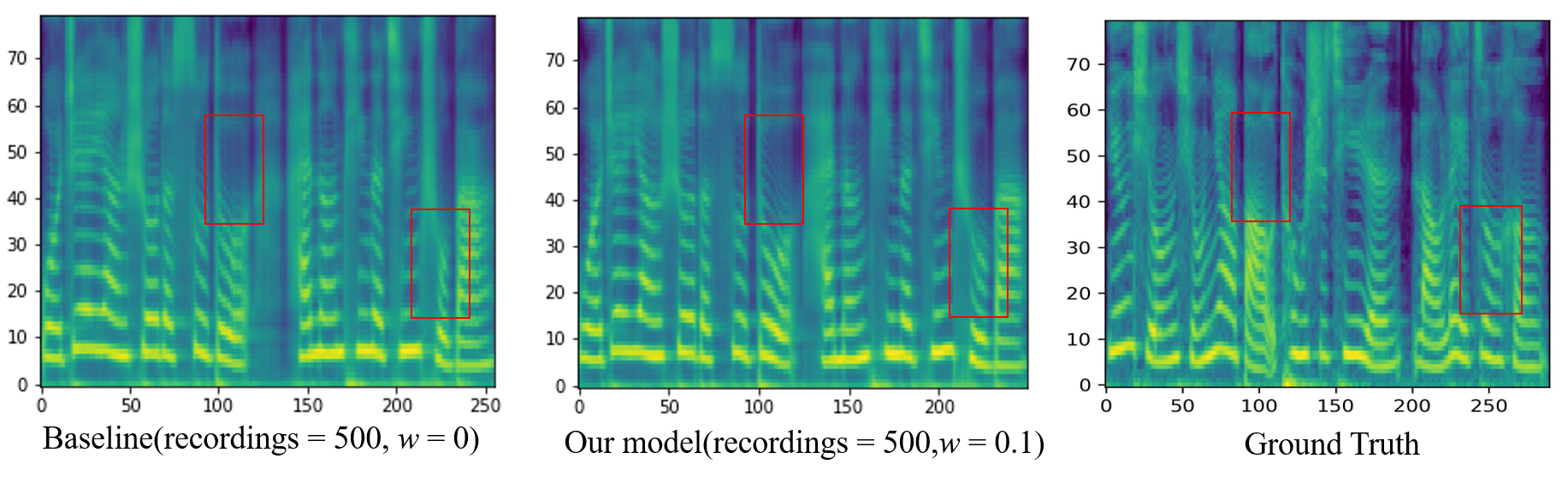}}
  \caption{Mel-spectrograms synthesized by baseline (recordings = 500, $\omega$ = 0)
and our model (recordings = 500, $\omega$ = 0.1) and the ground truth mel-spectrogram generated by the naural speech}
\label{fig:4}
\end{minipage}
\end{figure}

\begin{figure}[htb]
\begin{minipage}[b]{1.0\linewidth}
  \centering
  \centerline{\includegraphics[width=8cm]{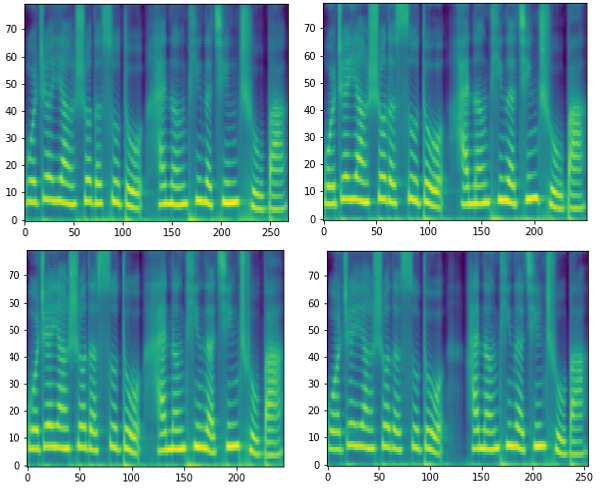}}
  \caption{Mel-spectrograms synthesized by our models use different size recordings.}
\label{fig:5}
\end{minipage}
\end{figure}

To further understand the reference loss's role. As shown in Figure~\ref{fig:4}, we visualize the synthesised mel-spectrugram with the groudtruth. Our model ($\omega$ = 0.1) fine is more apparent by observing parts 1 and 2 compared to the baseline ($\omega$ = 0); In the low frequency, we can find the speech's mel-spectrograms connection of the baseline is not smooth compared our model. As shown in Figure~\ref{fig:5}, we also visualize some mel-spectrograms of synthetic speech by our models which use different size recordings. It can be found from the results that the texture details of the high frequency and low-frequency parts of the mel-spectrogram under the condition of a small amount of sample data (recordings = 30, 300, 1000, 2000) can be well reflected, and they are both close to the best mel-spectrum our inferred by the fine-tuned model ($\omega = 0.1, recording = 2000$). 

From the results, we can find that the pseudo labels enlarge the numbers of the training samples during training. While the weight of the source model in the supervised model is controlled, that cloud does a trade-off between valid labels and pseudo labels. The improvement of quality compared with the baseline model also shows the training method has learned the label information for the unsupervised data samples.
\section{Conclusions}

To sum up, we proposed a method of reference model based. We studied the labeled target dataset overcomes the few-shot condition and the exposure bias problem. By implementing this scheme, we have verified the introduction of pseudo labels through the reference model to guide the fine-tuned model's training to achieve a regularization effect, which is a benefit to improve the adaptive ability of the original model.  What's more, it can also reduce the overfitting of the fine-tuned model during training on the limited target data. The result shows that the proposed method consistently outperforms the baseline systems under the few-shot condition.

\section{Acknowledgement}
This paper is supported by the Key Research and Development Program of Guangdong Province under grant No.2021B0101400003. Corresponding author is Jianzong Wang from Ping An Technology (Shenzhen) Co., Ltd (jzwang@188.com).

\bibliographystyle{IEEEtran}
\bibliography{mybib}

\begin{thebibliography}{10}
\providecommand{\url}[1]{#1}
\csname url@samestyle\endcsname
\providecommand{\newblock}{\relax}
\providecommand{\bibinfo}[2]{#2}
\providecommand{\BIBentrySTDinterwordspacing}{\spaceskip=0pt\relax}
\providecommand{\BIBentryALTinterwordstretchfactor}{4}
\providecommand{\BIBentryALTinterwordspacing}{\spaceskip=\fontdimen2\font plus
\BIBentryALTinterwordstretchfactor\fontdimen3\font minus
  \fontdimen4\font\relax}
\providecommand{\BIBforeignlanguage}[2]{{%
\expandafter\ifx\csname l@#1\endcsname\relax
\typeout{** WARNING: IEEEtran.bst: No hyphenation pattern has been}%
\typeout{** loaded for the language `#1'. Using the pattern for}%
\typeout{** the default language instead.}%
\else
\language=\csname l@#1\endcsname
\fi
#2}}
\providecommand{\BIBdecl}{\relax}
\BIBdecl

\bibitem{li2019neural}
N.~Li, S.~Liu, Y.~Liu, S.~Zhao, and M.~Liu, ``Neural speech synthesis with
  transformer network,'' in \emph{Proceedings of the AAAI Conference on
  Artificial Intelligence}, vol.~33, no.~01, 2019, pp. 6706--6713.

\bibitem{zeng2020aligntts}
Z.~Zeng, J.~Wang, N.~Cheng, T.~Xia, and J.~Xiao, ``Aligntts: Efficient
  feed-forward text-to-speech system without explicit alignment,'' in
  \emph{2020 IEEE International Conference on Acoustics, Speech and Signal
  Processing (ICASSP)}.\hskip 1em plus 0.5em minus 0.4em\relax IEEE, 2020, pp.
  6714--6718.

\bibitem{wang2019semantic}
C.~Wang, Y.~Wu, Y.~Du, J.~Li, S.~Liu, L.~Lu, S.~Ren, G.~Ye, S.~Zhao, and
  M.~Zhou, ``Semantic mask for transformer based end-to-end speech
  recognition,'' in \emph{21st Annual Conference of the International Speech
  Communication Association}.\hskip 1em plus 0.5em minus 0.4em\relax {ISCA},
  2020, pp. 971--975.

\bibitem{choi2020attentron}
S.~Choi, S.~Han, D.~Kim, and S.~Ha, ``Attentron: Few-shot text-to-speech
  utilizing attention-based variable-length embedding,'' \emph{Proc.
  Interspeech 2020}, pp. 2007--2011, 2020.

\bibitem{shen2018natural}
J.~Shen, R.~Pang, R.~J. Weiss, M.~Schuster, N.~Jaitly, Z.~Yang, Z.~Chen,
  Y.~Zhang, Y.~Wang, R.~Skerrv-Ryan \emph{et~al.}, ``Natural tts synthesis by
  conditioning wavenet on mel spectrogram predictions,'' in \emph{2018 IEEE
  International Conference on Acoustics, Speech and Signal Processing
  (ICASSP)}.\hskip 1em plus 0.5em minus 0.4em\relax IEEE, 2018, pp. 4779--4783.

\bibitem{wang2017tacotron}
Y.~Wang, R.~J. Skerry{-}Ryan, D.~Stanton, and et~al., ``Tacotron: Towards
  end-to-end speech synthesis,'' in \emph{18th Annual Conference of the
  International Speech Communication Association}.\hskip 1em plus 0.5em minus
  0.4em\relax {ISCA}, 2017, pp. 4006--4010.

\bibitem{zhao2022nnspeech}
B.~Zhao, X.~Zhang, J.~Wang, N.~Cheng, and J.~Xiao, ``nnspeech: Speaker-guided
  conditional variational autoencoder for zero-shot multi-speaker
  text-to-speech,'' in \emph{ICASSP2022}.\hskip 1em plus 0.5em minus
  0.4em\relax IEEE, 2022, pp. 4293--4297.

\bibitem{schmidt2019generalization}
F.~Schmidt, ``Generalization in generation: {A} closer look at exposure bias,''
  in \emph{Proceedings of the 3rd Workshop on Neural Generation and
  Translation}.\hskip 1em plus 0.5em minus 0.4em\relax Association for
  Computational Linguistics, 2019, pp. 157--167.

\bibitem{tang2022avqvc}
H.~Tang, X.~Zhang, J.~Wang, N.~Cheng, and J.~Xiao, ``Avqvc: One-shot voice
  conversion by vector quantization with applying contrastive learning,'' in
  \emph{ICASSP2022}.\hskip 1em plus 0.5em minus 0.4em\relax IEEE, 2022, pp.
  4613--4617.

\bibitem{bengio2015scheduled}
S.~Bengio, O.~Vinyals, N.~Jaitly, and N.~Shazeer, ``Scheduled sampling for
  sequence prediction with recurrent neural networks,'' in \emph{Advances in
  Neural Information Processing Systems}, 2015, pp. 1171--1179.

\bibitem{ranzato2015sequence}
M.~Ranzato, S.~Chopra, M.~Auli, and W.~Zaremba, ``Sequence level training with
  recurrent neural networks,'' in \emph{4th International Conference on
  Learning Representations}, 2016.

\bibitem{wang2022drvc}
Q.~Wang, X.~Zhang, J.~Wang, N.~Cheng, and J.~Xiao, ``Drvc: A framework of
  any-to-any voice conversion with self-supervised learning,'' in
  \emph{ICASSP2022}.\hskip 1em plus 0.5em minus 0.4em\relax IEEE, 2022, pp.
  3184--3188.

\bibitem{juang1985mixture}
B.-H. Juang and L.~Rabiner, ``Mixture autoregressive hidden markov models for
  speech signals,'' \emph{IEEE Transactions on Acoustics, Speech, and Signal
  Processing}, vol.~33, no.~6, pp. 1404--1413, 1985.

\bibitem{zhu2019pre}
X.~Zhu, Y.~Zhang, S.~Yang, L.~Xue, and L.~Xie, ``Pre-alignment guided attention
  for improving training efficiency and model stability in end-to-end speech
  synthesis,'' \emph{IEEE Access}, vol.~7, pp. 65\,955--65\,964, 2019.

\bibitem{zen2011product}
H.~Zen, M.~J. Gales, Y.~Nankaku, and K.~Tokuda, ``Product of experts for
  statistical parametric speech synthesis,'' \emph{IEEE Transactions on Audio,
  Speech, and Language Processing}, vol.~20, no.~3, pp. 794--805, 2011.

\bibitem{klatt1987review}
D.~H. Klatt, ``Review of text-to-speech conversion for english,'' \emph{The
  Journal of the Acoustical Society of America}, vol.~82, no.~3, pp. 737--793,
  1987.

\bibitem{asru2021zhang}
X.~Zhang, J.~Wang, N.~Cheng, E.~Xiao, and J.~Xiao, ``{CycleGEAN}:cycle
  generative enhanced adversarial network for voice conversion,'' in
  \emph{ASRU2021}.\hskip 1em plus 0.5em minus 0.4em\relax {IEEE}, 2021, pp.
  930--937.

\bibitem{tokuda2013speech}
K.~Tokuda, Y.~Nankaku, T.~Toda, H.~Zen, J.~Yamagishi, and K.~Oura, ``Speech
  synthesis based on hidden markov models,'' \emph{Proceedings of the IEEE},
  vol. 101, no.~5, pp. 1234--1252, 2013.

\bibitem{tokuda2000speech}
K.~Tokuda, T.~Yoshimura, T.~Masuko, T.~Kobayashi, and T.~Kitamura, ``Speech
  parameter generation algorithms for hmm-based speech synthesis,'' in
  \emph{{IEEE} International Conference on Acoustics, Speech, and Signal
  Processing}.\hskip 1em plus 0.5em minus 0.4em\relax {IEEE}, 2000, pp.
  1315--1318.

\bibitem{hunt1996unit}
A.~J. Hunt and A.~W. Black, ``Unit selection in a concatenative speech
  synthesis system using a large speech database,'' in \emph{Proceedings of the
  IEEE International Conference on Acoustics, Speech, and Signal Processing},
  vol.~1.\hskip 1em plus 0.5em minus 0.4em\relax IEEE, 1996, pp. 373--376.

\bibitem{black2007statistical}
A.~W. Black, H.~Zen, and K.~Tokuda, ``Statistical parametric speech
  synthesis,'' in \emph{Proceedings of the IEEE International Conference on
  Acoustics, Speech, and Signal Processing}, vol.~4.\hskip 1em plus 0.5em minus
  0.4em\relax IEEE, 2007, pp. IV--1229.

\bibitem{asru2021tang}
H.~Tang, X.~Zhang, J.~Wang, N.~Cheng, Z.~Zeng, E.~Xiao, and J.~Xiao, ``{TGAVC}:
  Improving autoencoder voice conversion with text-guided and adversarial
  training,'' in \emph{ASRU2021}.\hskip 1em plus 0.5em minus 0.4em\relax
  {IEEE}, 2021, pp. 938--945.

\bibitem{oord2016wavenet}
A.~van~den Oord, S.~Dieleman, H.~Zen, K.~Simonyan, O.~Vinyals, A.~Graves,
  N.~Kalchbrenner, A.~W. Senior, and K.~Kavukcuoglu, ``Wavenet: {A} generative
  model for raw audio,'' in \emph{The 9th {ISCA} Speech Synthesis Workshop},
  2016, p. 125.

\bibitem{griffin1984signal}
D.~Griffin and J.~Lim, ``Signal estimation from modified short-time fourier
  transform,'' \emph{IEEE Transactions on acoustics, speech, and signal
  processing}, vol.~32, no.~2, pp. 236--243, 1984.

\bibitem{chung2019semi}
Y.-A. Chung, Y.~Wang, W.-N. Hsu, Y.~Zhang, and R.~Skerry-Ryan,
  ``Semi-supervised training for improving data efficiency in end-to-end speech
  synthesis,'' in \emph{2019 IEEE International Conference on Acoustics, Speech
  and Signal Processing (ICASSP)}.\hskip 1em plus 0.5em minus 0.4em\relax IEEE,
  2019, pp. 6940--6944.

\bibitem{fei2006one}
L.~Fei-Fei, R.~Fergus, and P.~Perona, ``One-shot learning of object
  categories,'' \emph{IEEE transactions on pattern analysis and machine
  intelligence}, vol.~28, no.~4, pp. 594--611, 2006.

\bibitem{fink2005object}
M.~Fink, ``Object classification from a single example utilizing class
  relevance metrics,'' \emph{Advances in neural information processing
  systems}, vol.~17, pp. 449--456, 2005.

\bibitem{zhang2020adadurian}
Z.~Zhang, Q.~Tian, H.~Lu, L.-H. Chen, and S.~Liu, ``Adadurian: Few-shot
  adaptation for neural text-to-speech with durian,'' \emph{arXiv preprint
  arXiv:2005.05642}, 2020.

\bibitem{sharma2020strawnet}
M.~Sharma, T.~Kenter, and R.~Clark, ``Strawnet: Self-training wavenet for {TTS}
  in low-data regimes,'' in \emph{21st Annual Conference of the International
  Speech Communication Association}.\hskip 1em plus 0.5em minus 0.4em\relax
  {ISCA}, 2020, pp. 3550--3554.

\bibitem{moss2020boffin}
H.~B. Moss, V.~Aggarwal, N.~Prateek, J.~Gonz{\'a}lez, and R.~Barra-Chicote,
  ``Boffin tts: Few-shot speaker adaptation by bayesian optimization,'' in
  \emph{2020 IEEE International Conference on Acoustics, Speech and Signal
  Processing (ICASSP)}.\hskip 1em plus 0.5em minus 0.4em\relax IEEE, 2020, pp.
  7639--7643.

\bibitem{chen2018sample}
Y.~Chen, Y.~M. Assael, B.~Shillingford, D.~Budden, S.~E. Reed, H.~Zen, Q.~Wang,
  L.~C. Cobo, A.~Trask, B.~Laurie, {\c{C}}.~G{\"{u}}l{\c{c}}ehre, A.~van~den
  Oord, O.~Vinyals, and N.~de~Freitas, ``Sample efficient adaptive
  text-to-speech,'' in \emph{7th International Conference on Learning
  Representations}, 2019.

\bibitem{bollepalli2019lombard}
B.~Bollepalli, L.~Juvela, P.~Alku \emph{et~al.}, ``Lombard speech synthesis
  using transfer learning in a tacotron text-to-speech system.'' in \emph{20th
  Annual Conference of the International Speech Communication Association},
  2019, pp. 2833--2837.

\bibitem{deng2018modeling}
Y.~Deng, L.~He, and F.~K. Soong, ``Modeling multi-speaker latent space to
  improve neural {TTS:} quick enrolling new speaker and enhancing premium
  voice,'' \emph{CoRR}, vol. abs/1812.05253, 2018.

\bibitem{hinton2015distilling}
G.~Hinton, O.~Vinyals, and J.~Dean, ``Distilling the knowledge in a neural
  network,'' \emph{arXiv preprint arXiv:1503.02531}, 2015.

\bibitem{liu2020teacher}
R.~Liu, B.~Sisman, J.~Li, F.~Bao, G.~Gao, and H.~Li, ``Teacher-student training
  for robust tacotron-based tts,'' in \emph{ICASSP 2020-2020 IEEE international
  conference on acoustics, speech and signal processing (ICASSP)}.\hskip 1em
  plus 0.5em minus 0.4em\relax IEEE, 2020, pp. 6274--6278.

\bibitem{lee2013pseudo}
D.-H. Lee \emph{et~al.}, ``Pseudo-label: The simple and efficient
  semi-supervised learning method for deep neural networks,'' in \emph{Workshop
  on challenges in representation learning, ICML}, vol.~3, no.~2, 2013, p. 896.

\bibitem{higuchi2021advancing}
Y.~Higuchi, N.~Moritz, J.~L. Roux, and T.~Hori, ``Advancing momentum
  pseudo-labeling with conformer and initialization strategy,'' \emph{arXiv
  preprint arXiv:2110.04948}, 2021.

\bibitem{ren2020fastspeech}
Y.~Ren, C.~Hu, X.~Tan, T.~Qin, S.~Zhao, Z.~Zhao, and T.-Y. Liu, ``Fastspeech 2:
  Fast and high-quality end-to-end text to speech,'' \emph{arXiv preprint
  arXiv:2006.04558}, 2020.

\bibitem{kong2020hifi}
J.~Kong, J.~Kim, and J.~Bae, ``Hifi-gan: Generative adversarial networks for
  efficient and high fidelity speech synthesis,'' \emph{Advances in Neural
  Information Processing Systems}, vol.~33, pp. 17\,022--17\,033, 2020.

\bibitem{xie2020self}
Q.~Xie, M.-T. Luong, E.~Hovy, and Q.~V. Le, ``Self-training with noisy student
  improves imagenet classification,'' in \emph{2020 {IEEE/CVF} Conference on
  Computer Vision and Pattern Recognition}, 2020, pp. 10\,687--10\,698.

\end{thebibliography}
\end{document}